\documentclass[
showkeys,12pt,
preprint,preprintnumbers,nofootinbib,
groupedaddress,superscriptaddress,amsmath,amssymb]{revtex4}
\usepackage{graphicx}
\usepackage{dcolumn}
\usepackage{bm}
\usepackage{amssymb}
\usepackage{amsmath}
\usepackage{epsfig}    
\usepackage{color}
\usepackage{slashed}
\usepackage{hhline}

\def\be{\begin{equation}}
\def\ee{\end{equation}}
\newcommand{\bea}{\begin{eqnarray}}
\newcommand{\eea}{\end{eqnarray}}
\newcommand{\nn}{\nonumber}

\numberwithin{equation}{section}

\begin{document}

{\begin{flushright}{KIAS-P16004}
\end{flushright}}

\title{Four-loop Neutrino Model \\
Inspired by Diphoton Excess at 750 GeV}
%
\author{Takaaki Nomura}
\email{nomura@kias.re.kr}
\affiliation{School of Physics, KIAS, Seoul 130-722, Korea}

\author{Hiroshi Okada}
\email{macokada3hiroshi@gmail.com}
\affiliation{Physics Division, National Center for Theoretical Sciences, Hsinchu, Taiwan 300}

\date{\today}

\begin{abstract}
We propose a four-loop induced radiative neutrino mass model inspired by the diphoton excess at 750 GeV recently reported by ATLAS and CMS, in which a sizable diphoton excess is obtained via photon fusion introducing multi doubly-charged scalar bosons. Also we discuss the muon anomalous magnetic moment, and a dark matter candidate. The main process to explain the observed relic density relies on the final state of the new particle at 750 GeV. Finally we show the numerical results and obtain allowed region of several physical values in our model.
\end{abstract}
\maketitle
\newpage

\section{Introduction}
\begin{figure}[tb]
\begin{center}
\includegraphics[scale=0.6]{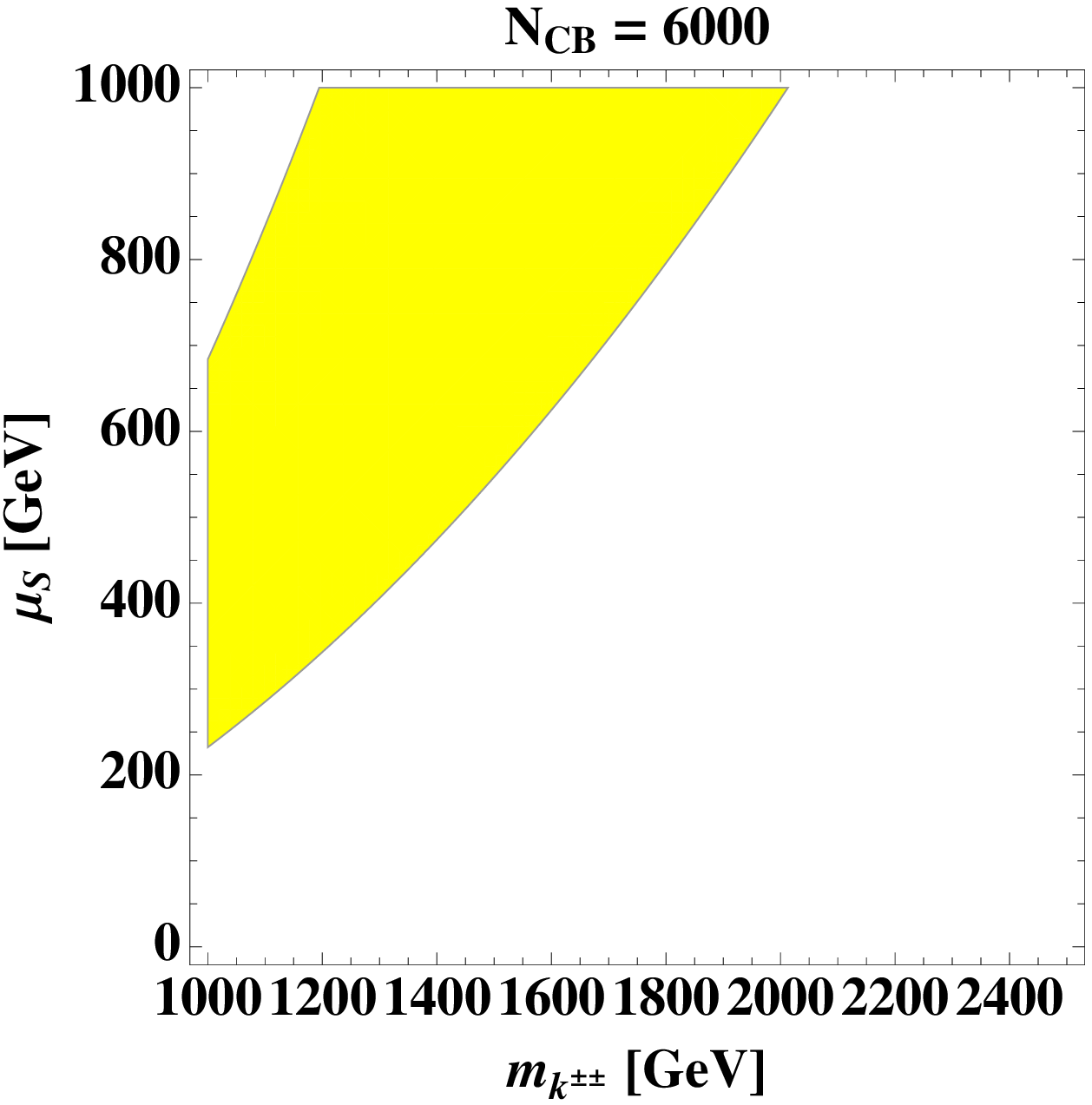}
\includegraphics[scale=0.6]{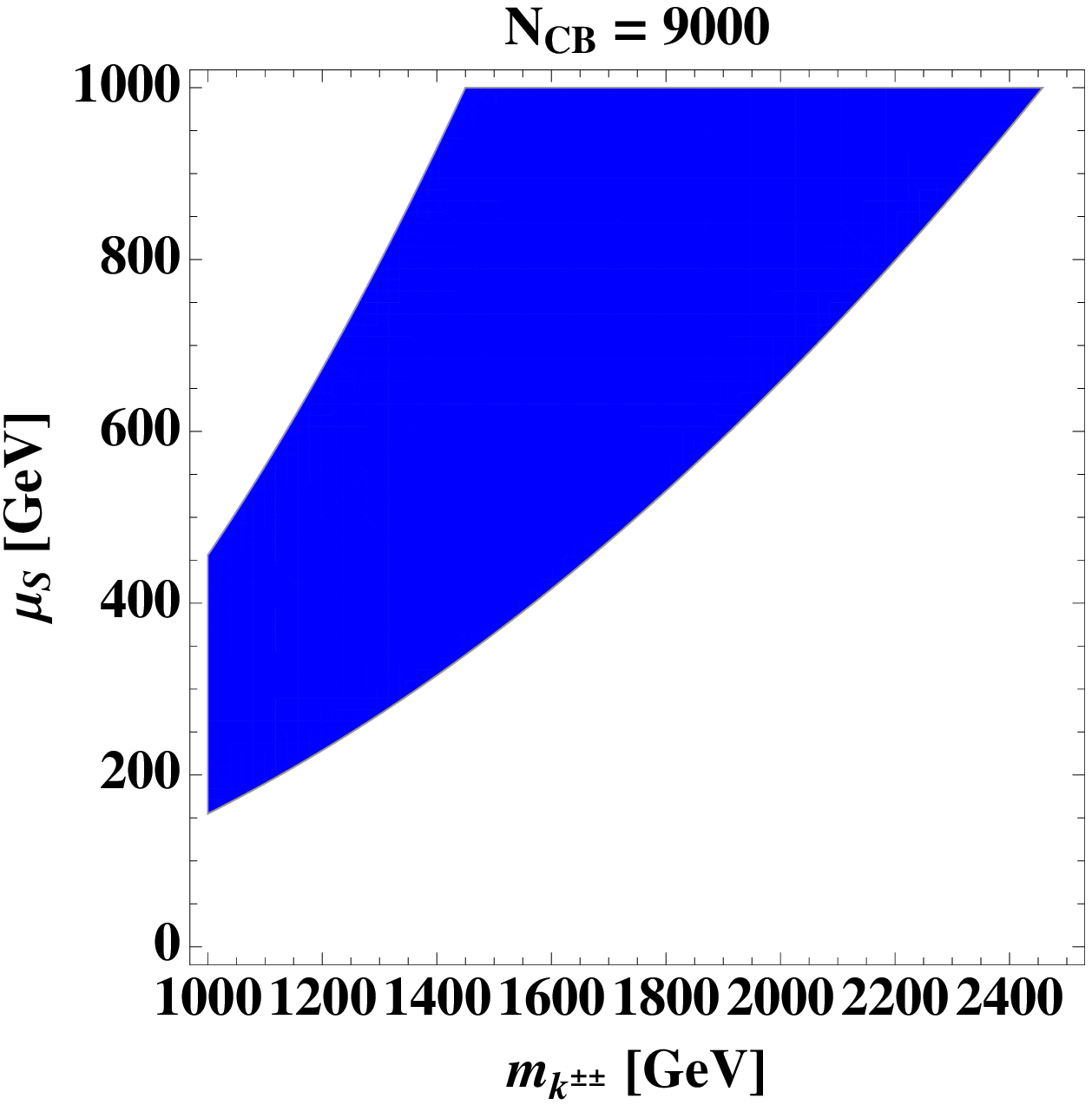}
\caption{ The figures represent the allowed region between the mass of $k^{\pm\pm}$ and the trilinear term of $\mu_S$ to satisfy
$5.3\ {\rm GeV}\lesssim\Gamma_S\lesssim 45$ GeV coming from the experimental resolution and  the best fit value of ATLAS respectively, where each of colored region (yellow for $N_{CB}=6000$ and blue for $N_{CB}=9000$) is allowed only and the upper line corresponds to $\Gamma_S=45$ GeV and the lower line corresponds to $\Gamma_S=5.3$ GeV.
}   \label{mk-ncb}
\end{center}\end{figure}

According to the recent announcements by ATLAS and CMS experiments, a new particle could exist at around 750 GeV  by the observation of the diphoton invariant mass spectrum from the run-II data in 13 TeV~\cite{ATLAS-CONF-2015-081,CMS:2015dxe}.
Subsequently a vast of paper along this line of issue has been arisen in Ref.~\cite{Harigaya:2015ezk, Mambrini:2015wyu,Backovic:2015fnp,Angelescu:2015uiz,Nakai:2015ptz,Knapen:2015dap,Buttazzo:2015txu,Pilaftsis:2015ycr,Franceschini:2015kwy,DiChiara:2015vdm,Higaki:2015jag,McDermott:2015sck,Ellis:2015oso,Low:2015qep,Bellazzini:2015nxw,Gupta:2015zzs,Petersson:2015mkr,Molinaro:2015cwg,Dutta:2015wqh,Cao:2015pto,Matsuzaki:2015che,Kobakhidze:2015ldh,Martinez:2015kmn,Cox:2015ckc,Becirevic:2015fmu,No:2015bsn,Demidov:2015zqn,Chao:2015ttq,Fichet:2015vvy,Curtin:2015jcv,Bian:2015kjt,Chakrabortty:2015hff,Ahmed:2015uqt,Agrawal:2015dbf,Csaki:2015vek,Falkowski:2015swt,Aloni:2015mxa,Bai:2015nbs,Gabrielli:2015dhk,Benbrik:2015fyz,Kim:2015ron,Alves:2015jgx,Megias:2015ory,Carpenter:2015ucu,Bernon:2015abk,Chao:2015nsm,Arun:2015ubr,Han:2015cty,Chang:2015bzc,Chakraborty:2015jvs,Ding:2015rxx,Han:2015dlp,Han:2015qqj,Luo:2015yio,Chang:2015sdy,Bardhan:2015hcr,Feng:2015wil,Antipin:2015kgh,Wang:2015kuj,Cao:2015twy,Huang:2015evq,Liao:2015tow,Heckman:2015kqk,Dhuria:2015ufo,Bi:2015uqd,Kim:2015ksf,Berthier:2015vbb,Cho:2015nxy,Cline:2015msi,Bauer:2015boy,Chala:2015cev,Barducci:2015gtd,Boucenna:2015pav,Murphy:2015kag,Hernandez:2015ywg,Dey:2015bur,Pelaggi:2015knk,deBlas:2015hlv,Belyaev:2015hgo,Dev:2015isx,Huang:2015rkj,Moretti:2015pbj,Patel:2015ulo,Badziak:2015zez,Chakraborty:2015gyj,Cao:2015xjz,Altmannshofer:2015xfo,Cvetic:2015vit,Gu:2015lxj,Allanach:2015ixl,Davoudiasl:2015cuo,Craig:2015lra,Das:2015enc,Cheung:2015cug,Liu:2015yec,Zhang:2015uuo,Casas:2015blx,Hall:2015xds,Han:2015yjk,Park:2015ysf,Salvio:2015jgu,Chway:2015lzg,Li:2015jwd,Son:2015vfl,Tang:2015eko,An:2015cgp,Cao:2015apa,Wang:2015omi,Cai:2015hzc,Cao:2015scs,Kim:2015xyn,Gao:2015igz,Chao:2015nac,Bi:2015lcf,Goertz:2015nkp,Anchordoqui:2015jxc,Dev:2015vjd,Bizot:2015qqo,Ibanez:2015uok,Chiang:2015tqz,Kang:2015roj,Hamada:2015skp,Huang:2015svl,Kanemura:2015bli,Kanemura:2015vcb,Low:2015qho,Hernandez:2015hrt,Jiang:2015oms,Kaneta:2015qpf,Marzola:2015xbh,Ma:2015xmf,Dasgupta:2015pbr}.
One of these interpretations is to identify a scalar (or pseudoscalar) as the new particle ($S$), and the resonance occurs in the process;
$pp\to S+{ X}\to 2\gamma+{X}$, where $X$ is the missing particle. This can be interpreted as the following 13 TeV data in terms of the production cross section of $S$ and its branching ratio of two photons,
\begin{align}
&\mu_{\rm ATLAS}=\sigma(2p\to S+X)\times {BR}(S\to 2\gamma)=(6.2^{+2.4}_{-2.0})\ {\rm fb},\\
&\mu_{\rm CMS}=\sigma(2p\to S+X)\times {BR}(S\to 2\gamma)=(5.6\pm{2.4})\ {\rm fb},
\label{lhc-exp}
\end{align}
which is extremely large compared to the previous observations from the run-I data at 8 TeV~\cite{Aad:2015mna, CMS:2014onr}. 
Also the ATLAS experiment group~\cite{ATLAS-CONF-2015-081} reported $\Gamma_S=45$ GeV that is the best fit value of the decay width of $S$ to the two photons, and $\Gamma_S=5.3$ GeV is given as the experimental resolution obtained by the analysis~\cite{McDermott:2015sck}.
To achieve such a large signal strength, we have to enlarge the production cross section and (or) its branching ratio.
One of the simplest ways to enhance the production cross section is to introduce a vector like exotic quark that couples to $S$,
where such a quark induces the gluon fusion production of $S$ that can be always dominant process~\cite{Franceschini:2015kwy}. 
On the other hand, one of the simplest  ways to increase the branching ratio to photons is that $S$ should couples to the isospin singlet bosons or fermions with nonzero electric charges,
because main modes such as a pair of $W^\pm$ bosons can be forbidden.
However once one can reach the enough branching ratio to the two photons, (which is around $\approx$ 60 \%), the dominant production cross section can also be arisen from the photon fusion process, which is proposed by, {\it i.e.}, Ref.~\cite{Csaki:2015vek}.
This scenario is in favor of leptonic models, especially, radiative seesaw models, when such charged particles  also interact with lepton sector. 
{Especially there are some representative radiative seesaw models at the three-loop level~\cite{Krauss:2002px, Aoki:2008av, Gustafsson:2012vj}.}
In this framework, the recent paper~\cite{Kanemura:2015bli} has concluded that
{\it the ${\cal O}(10^3-10^4$) number of electrically charged bosons that propagate between $S$ and two photons have to be introduced} as can be seen in Fig.~\ref{mk-ncb},~\footnote{The diphoton excess is analyzed by rather general way, introducing arbitral number of doubly charged bosons with isospin singlet in this paper, although they fix a specific model in the neutrino sector. Hence one can apply some results to any kind of leptonic models that include charged bosons with isospin singlet even when singly charged bosons.} in order to satisfy the condition of unitarity bound via processes such as $k^{\pm\pm} S\to k^{\pm\pm} \to k^{\pm\pm} S$ and $2k^{\pm\pm} \to S \to 2k^{\pm\pm}$.
Therefore, the trilinear term $\mu_S$ proportional to $Sk^{\pm\pm}k^{\mp\mp}$ should be nearly equal or less than $m_S\approx$ 750 GeV.  
The relevant potential per $k^{\pm\pm}$  to generate the diphoton anomaly is simply given by
\begin{align}
{\cal V} = \mu_S S k^{++}k^{--} + m_{k}  k^{++}   k^{--}  +{\rm c.c.}.
\end{align}
 Then the  total cross section with $m_S=$750 GeV at 13 TeV is given by~\cite{Csaki:2015vek}
\begin{align}
\sigma_{\gamma\gamma}(\equiv \sigma(2p\to 2\gamma+X))= \left(\frac{\Gamma_S}{45\ {\rm GeV}}\right)\times BR^2(S\to 2\gamma)\times (73-162)\ {\rm fb}.
\end{align}
In our case the cross section simplifies the following values due to $BR(S\to 2\gamma)\approx60$\%, 
\begin{align}
\left(3.0\ {\rm fb}\lesssim \sigma_{\gamma\gamma}(\Gamma_S=5.3 {\rm GeV}) \lesssim 6.7\ {\rm fb}\right) 
-
\left(25.5\ {\rm fb}\lesssim \sigma_{\gamma\gamma}(\Gamma_S=45 {\rm GeV})\lesssim 56.6\ {\rm fb}\right),
\label{eq:dip}
\end{align}
that satisfies the data in Eq.~(\ref{lhc-exp}). 
Here we use the value $5.3\ {\rm GeV}\lesssim\Gamma_S\lesssim 45$ GeV coming from the best fit value of ATLAS and the experimental resolution, and we find allowed regions in terms of $m_{k^{\pm\pm}}$ and $\mu_S$ to satisfy the decay width depending on the number of charged bosons $N_{CB}$ as can be seen in Fig.~\ref{mk-ncb}.

This result could drastically changes the situation of any radiative seesaw models that include electrically charged bosons such as Zee-Babu model~\cite{zee-babu}, which is the first proposal including the doubly charged boson, because the scale of neutrino masses must be enhanced by $N_{CB}$.
To show this issue more clearly, let us consider the Zee-Babu model. 
The model has the following relevant terms per $k^{\pm\pm}$:
{\begin{align}
-{\cal L}
&\supset  y_\ell \ell_L \Phi e_R +  f \bar\ell^c_L \ell_L {h^+}+ g \bar e_R^c e_R k^{++} +
\mu h^+ h^+ k^{--} + {\rm c.c.}.
\end{align}
Then the resulting neutrino mass has to be multiplied by $N_{CB}$, and can be estimated as 
\begin{align}
m_\nu\approx \frac{16 N_{CB}}{(4\pi)^4} \frac{\mu g^*(f m_\ell)^2}{M_{\rm max}^2} \times ({\rm loop\ factor)}
& \lesssim
\frac{{\cal O}(10^3-10^4)}{16\pi^4} \frac{\mu g^* f^2[{\rm GeV}]^2}{M_{\rm max}^2}\nn\\
&\approx {\cal O}(1-10) \frac{\mu g^* f^2[{\rm GeV}]^2}{M_{\rm max}^2},
\end{align}
where $M_{\rm max}\equiv{\rm Max}[m_{k^{\pm\pm}},m_{h^\pm}]$,}
we have used $m_\ell=m_\tau\approx {\cal O}$(1) [GeV], and loop factor is order 1.
It suggests that the neutrino mass scale is determined by the trilinear coupling $\mu$ and the Yukawa couplings,
and $N_{CB}$ that almost compensates the two loop suppression effect.
Therefore the two loop neutrino mass scale is equivalent to the tree level scale.
Applying this fact, we will discuss our radiative neutrino model at the four loop level in the next section, which could be equivalent to a typical
 two loop radiative model. Then we will conclude and discuss in Sec.~III.


\newpage

\section{ Model setup and Analysis}
 \begin{widetext}
\begin{center} 
\begin{table}[tbc]
\begin{tabular}{|c||c|c|c|c||c|c|c|c|}\hline\hline  
&\multicolumn{4}{c||}{Lepton Fields} & \multicolumn{4}{c|}{Scalar Fields} \\\hline
& ~$L_L$~ & ~$e_R^{}$~ & ~$E^{}_{}$ ~ & ~$N_R$~  & ~$\Phi$~ & ~$S$  & ~$h^+$  & ~$k^{++}$ \\\hline 
$SU(2)_L$ & $\bm{2}$  & $\bm{1}$  & $\bm{1}$ & $\bm{1}$ & $\bm{2}$ & $\bm{1}$  & $\bm{1}$   & $\bm{1}$ \\\hline 
$U(1)_Y$ & $-1$ & $-1$  & $-2$ & $0$ & $0$ & $0$ & $1$  & $2$  \\\hline
$U(1)$ & $\ell$ & $\ell$  & $3\ell$ & $\frac{\ell}3$ & $0$ & $\frac{2\ell}3$ & $-2\ell$  & $-\frac{10\ell}3$  \\\hline
\end{tabular}
\caption{Contents of fermion and scalar fields
and their charge assignments under $SU(2)_L\times U(1)_Y\times U(1)$.}
\label{tab:1}
\end{table}
\end{center}
\end{widetext}

In this section, we explain our model with global $U(1)$ symmetry. 
The particle contents and their charges are shown in Tab.~\ref{tab:1}.
We add a vector-like exotic doubly charged fermion $E$, a Majorana fermion $N_R$,
a singly charged scalar $h^\pm$,  {\it the $N_{CB}$ number of doubly charged scalars $k^{\pm\pm}$}, and a neutral scalar $S$ to the SM, where all these new fields are iso-spin singlet, and $S$ is identified as a new scalar with 750 GeV mass.
We assume that  only the SM Higgs $\Phi$  and $S$ have vacuum
expectation values (VEVs), which are respectively symbolized by $v/\sqrt2$ and $v_S/\sqrt2$. 
The quantum number $\ell\neq0$ of $U(1)$ symmetry is arbitrary, but its assignment for each field is unique to realize our four loop neutrino model. 

The relevant Lagrangian and Higgs potential under these symmetries  per $k^{\pm\pm}$ are given by
\begin{align}
-\mathcal{L}_{Y}
&\supset
y_{\ell} \bar L_{L} \Phi e_{R}+
{ f_{} \bar L_{L}^c i \tau_2 L_{L} h^+ }
+g_{} \bar E_{L} e_R  h^{-}  +h \bar N_{R} E^c_R  k^{--}   + \frac{y_N }{2} S^*\bar N^c_R N_R + { M_{E}} \bar E_{L} E_{R} \nn\\
&
-  \lambda_{hk} S^* h^-h^- k^{++} -  \lambda_{Sk} |S|^2 k^{++}k^{--} 
+ {\rm h.c.},
\label{Eq:lag-flavor}
\end{align}
{where $\tau_2$ is a second component of the Pauli matrix.}
After the global $U(1)$ spontaneous breaking of $S$, we obtain trilinear terms as well as the Majorana masses as follows: 
\begin{align}
-\mathcal{L}_{Y}
&\supset 
 \frac{M_N}{2} \bar N^c_R N_R 
- \mu h^-h^- k^{++} - \mu_S Sk^{++}k^{--} 
+ {\rm h.c.},
\label{Eq:lag-flavor}
\end{align}
where $M_N\equiv y_N v_S/\sqrt2$, $\mu\equiv \lambda_{hk}v_S/\sqrt2$, and $\mu_S\equiv \lambda_{Sk}v_S/\sqrt2$. 
The first term of $\mathcal{L}_{Y}$ generates the SM
charged-lepton masses $m_\ell\equiv y_\ell v/\sqrt2$ after the electroweak spontaneous breaking of $\Phi$.
We work on the basis where all the coefficients are real and positive for simplicity. 
The isospin doublet scalar field can be parameterized as $\Phi=[w^+,\frac{v+\phi+iz}{\sqrt2}]^T$
where $v~\simeq 246$ GeV is VEV of the Higgs doublet, and $w^\pm$
and $z$ are respectively absorbed by the longitudinal component of $W$ and $Z$ boson.
The isospin singlet scalar field can be parameterized as $S=\frac{v_S+ s}{\sqrt2}e^{iG/v_S}$.
Here we assume $\phi$ is the SM Higgs, therefore we neglect the mixing between $\phi$ and $s$ for simplicity.
We also assume that the lightest Majorana fermion $N_R|_{\rm lightest} = X$ does not couple to $E_R$ and $k^{\pm \pm}$ in the fourth term of ${\cal L}_Y$ and does not mix with other $N_R$
so that it can be stable and a DM candidate.
Such a situation for DM can easily be realized by imposing additional $Z_2$ odd assignment.

\begin{figure}[tb]
\begin{center}
\includegraphics[clip, width=60mm,angle=-90]{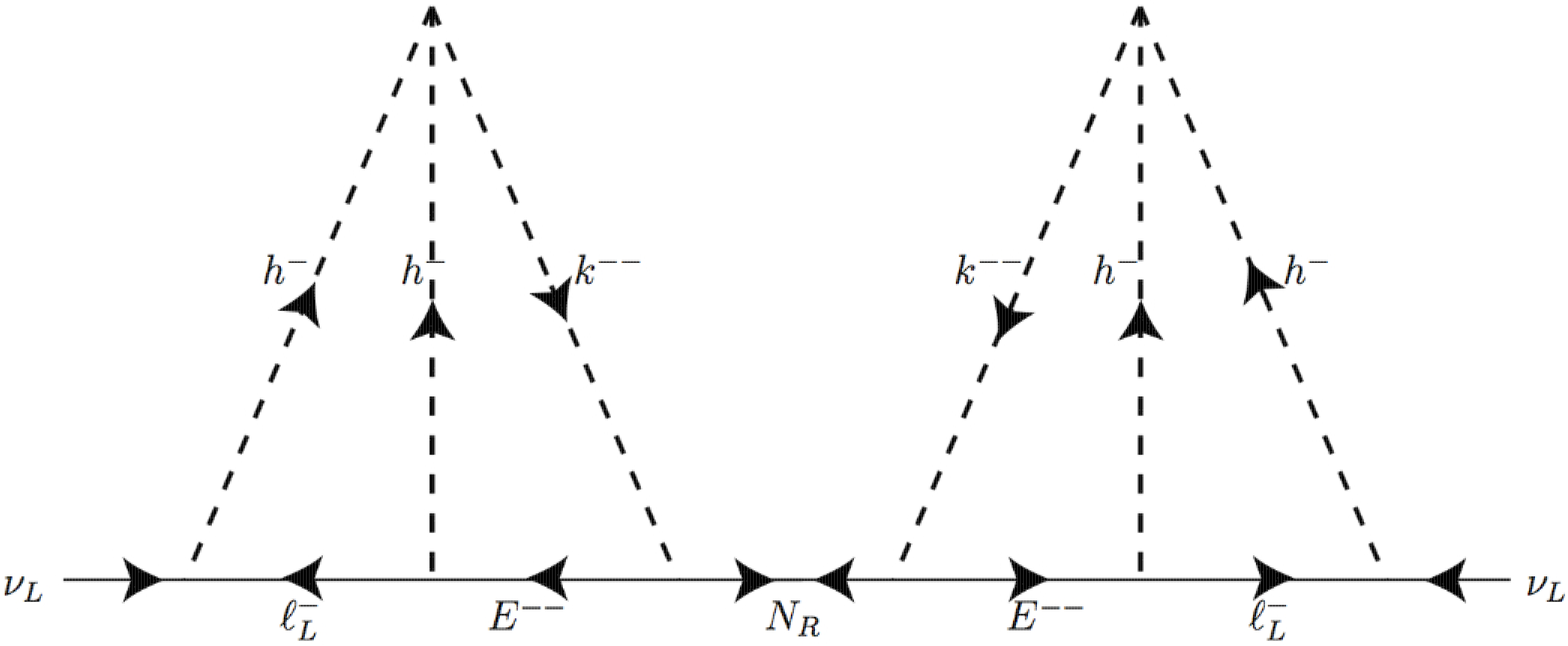}
\includegraphics[clip, width=60mm,angle=-90]{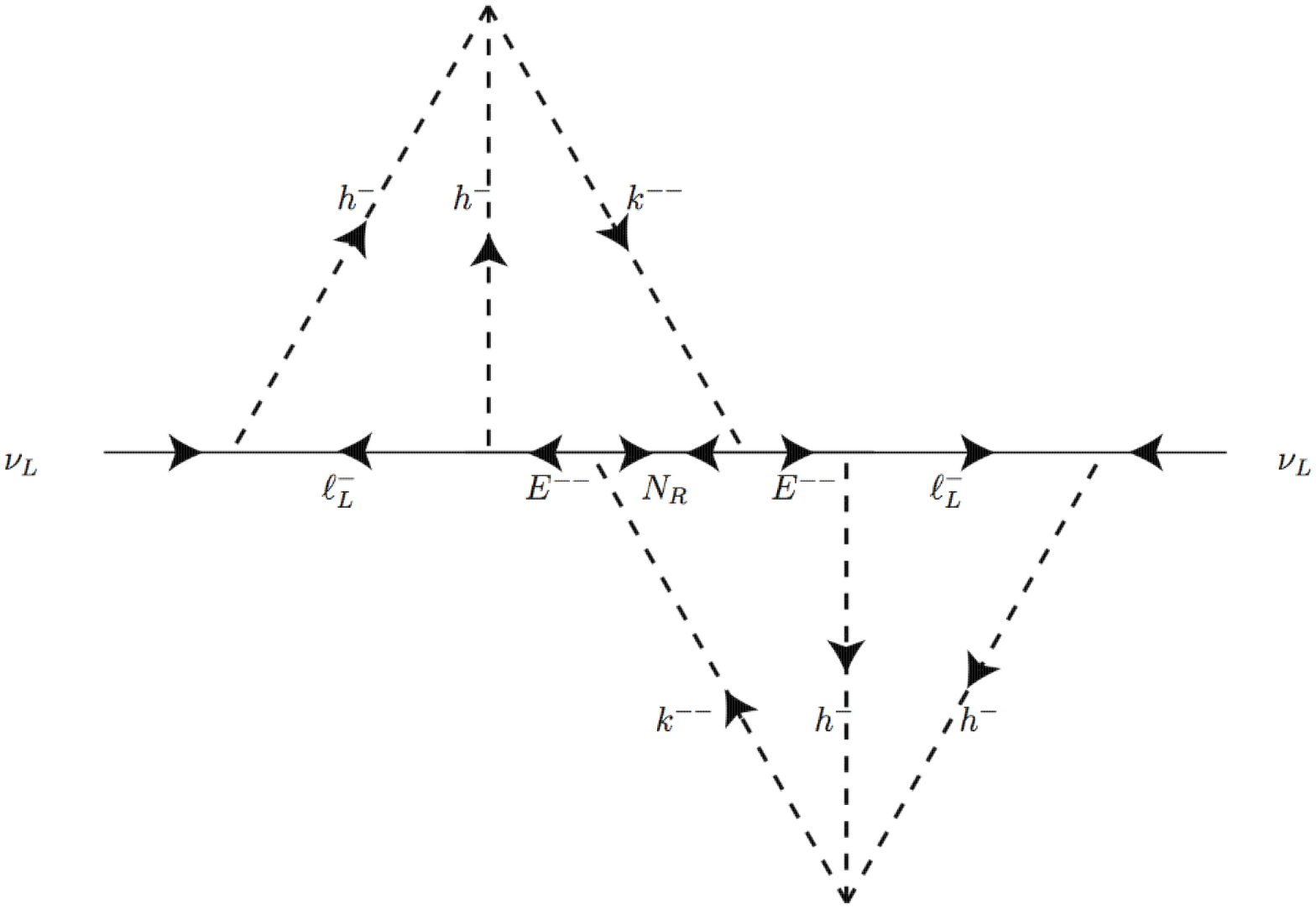}
\includegraphics[clip, width=60mm,angle=-90]{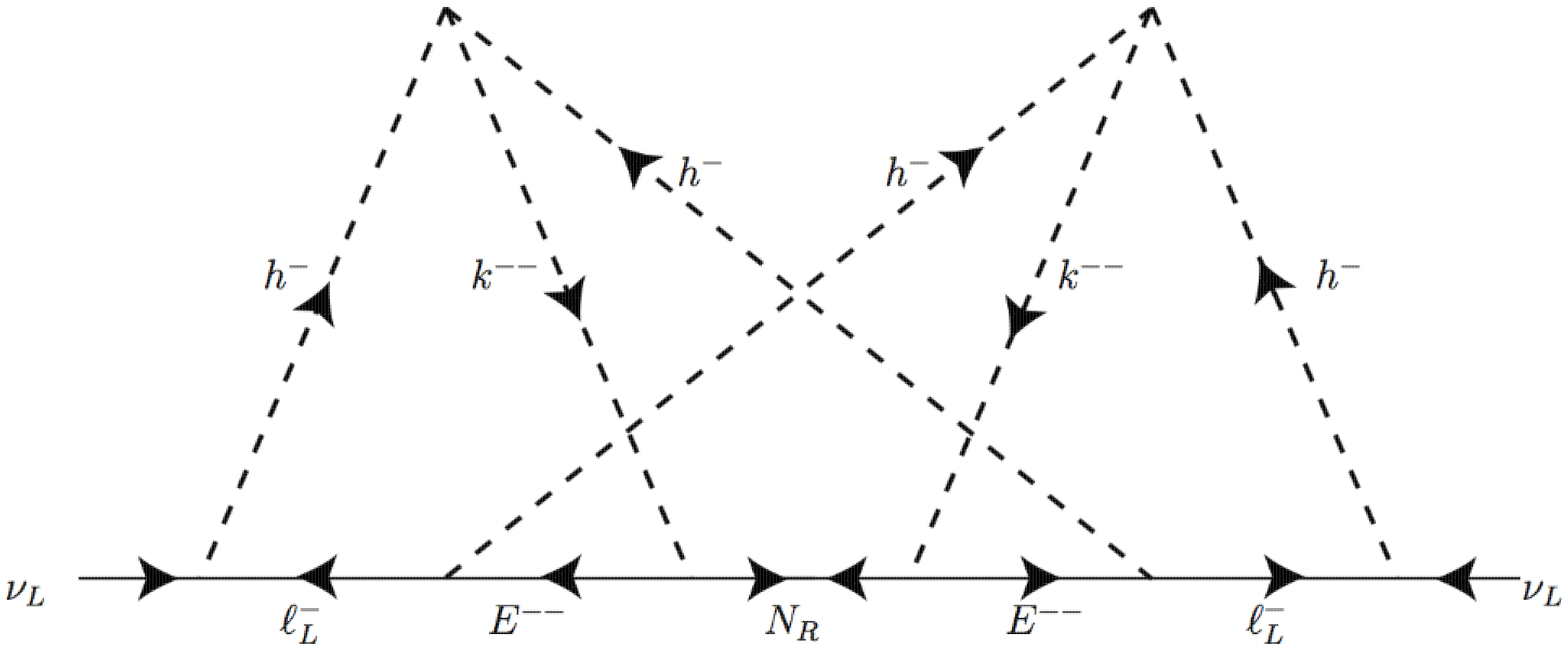}
\includegraphics[clip, width=60mm,angle=-90]{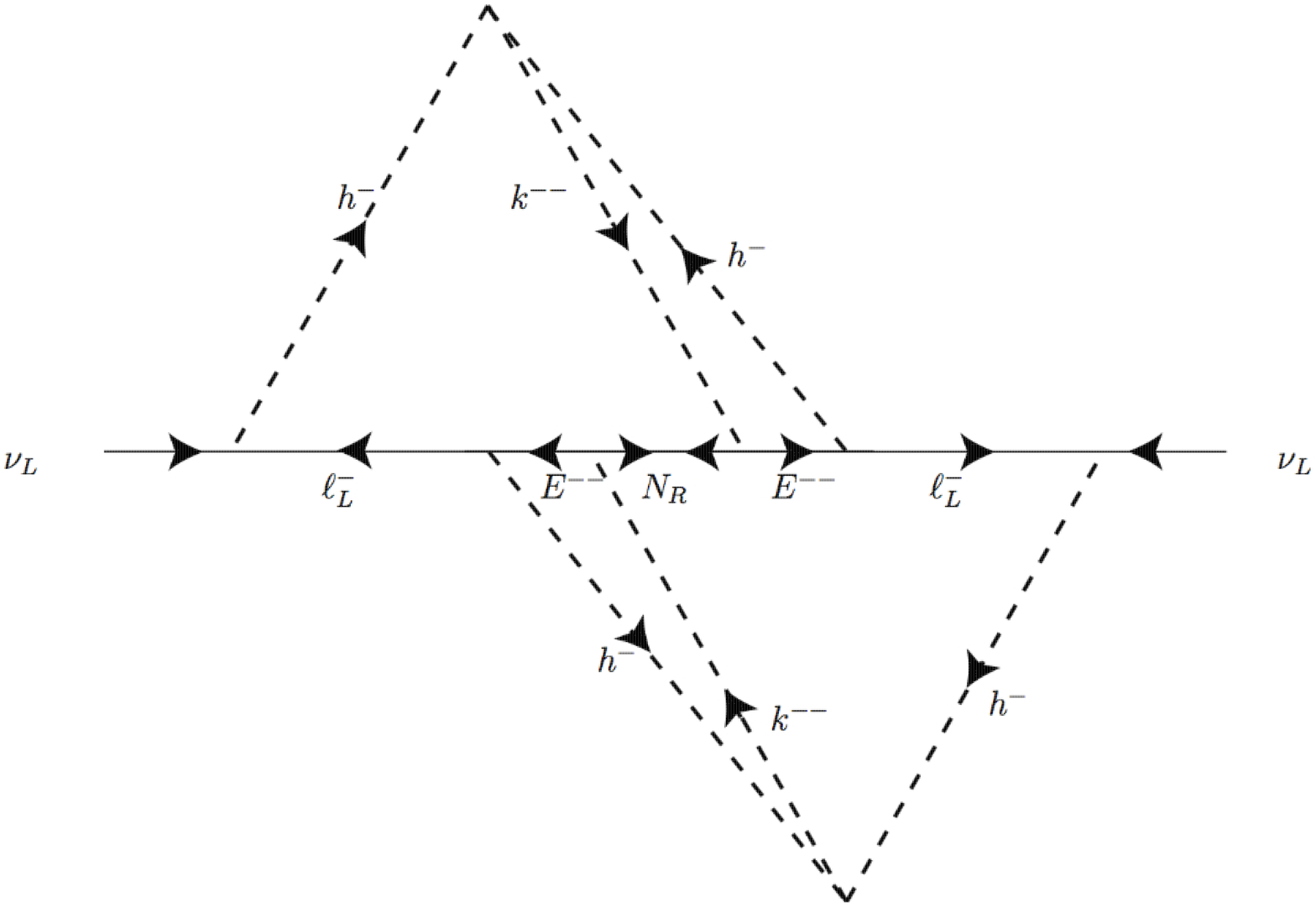}
\caption{ Neutrino masses at the one-loop level.
}   \label{fig:neut1}
\end{center}\end{figure}
{\it Neutrino mass matrix}:

Then the leading contribution to the active neutrino masses $m_\nu$  is given at four-loop level as shown in Figure~\ref{fig:neut1}, and we can respectively estimate the order of masses as follows:
\begin{align}
m_{\nu}&=m_{\nu}^{I}+m_{\nu}^{II}+m_{\nu}^{III}+m_{\nu}^{IV},\\
m_{\nu}^{I}&\approx \frac{[N_{CB} \mu m_\ell M_E f g^* h]^2}{(4\pi)^8 M^4_{\rm max} M_N} G_I(x_\ell,x_E,x_h,x_k)
,\label{mnu1}\\
m_{\nu}^{i}&\approx \frac{N_{CB}^2 M_R [\mu m_\ell M_E f g^* h ]^2}{(4\pi)^8 M^6_{\rm max} } { G}_i(x_\ell,x_E,x_h,x_k,x_{N}),\quad {(i=II-IV)},
\label{mnu2}
\end{align}
where the left-top side of figure corresponds to $m_{\nu}^{I}$,  the right-top side of figure corresponds to $m_{\nu}^{II}$, 
 the left-bottom side of figure corresponds to $m_{\nu}^{III}$,  the right-bottom side of figure corresponds to $m_{\nu}^{IV}$,
and we define $x_{i}\equiv (m_i/M_{\rm max})^2$ and $M_{\rm max}={\rm Max}[M_E,m_h,m_k,M_N]$. $G_I$ consists of two pairs of the Zee-Babu like two-loop function.
Obviously $m_\nu^I$ can be greater than $m_\nu^{II-IV}$ under the condition $G_I\approx{G_i}$, since the ratio is given by
\begin{align}
\frac{m_\nu^I}{m_\nu^{II-IV}}\approx \left(\frac{M_{\rm max}}{M_N}\right)^2 \gg 1.
\end{align}
Hence we can approximate the neutrino masses as 
\begin{align}
m_{\nu}&=m_{\nu}^{I}\approx \frac{[N_{CB} \mu m_\ell M_E f g^* h]^2}{(4\pi)^8 M^4_{\rm max} M_N},
\end{align}
where  we take $G_I ={\cal O}$(1), and $m_\nu$ should be $0.001\ {\rm eV}\lesssim m_\nu \lesssim 0.1\ {\rm eV}$
from the neutrino oscillation data~\cite{pdf}.

{\it Muon anomalous magnetic moment}:

The muon anomalous magnetic moment (muon $g-2$) has been 
measured at Brookhaven National Laboratory
that suggests there is a discrepancy between the
experimental data and the prediction in the SM. 
The difference $\Delta a_{\mu}\equiv a^{\rm exp}_{\mu}-a^{\rm SM}_{\mu}$
is respectively calculated in Ref.~\cite{discrepancy1} and Ref.~\cite{discrepancy2} as 
\begin{align}
\Delta a_{\mu}=(29.0 \pm 9.0)\times 10^{-10},\
\Delta a_{\mu}=(33.5 \pm 8.2)\times 10^{-10}. \label{dev12}
\end{align}
The above results given in Eq. (\ref{dev12}) correspond
to $3.2\sigma$ and $4.1\sigma$ deviations, respectively. 
Our formula of muon $g-2$ is given by
\begin{align}
&\Delta a_\mu\approx \frac{N_{CB} m_\mu^2}{(4\pi)^2}\left[(g^*g)_{22} F(E,h)-\frac{(f^*f)_{22}}{3 m_{h^\pm}^2}\right],\label{eq:g-2}
\\
&F(E,h)\approx \frac{4 M_E^6-9M_E^4 m^2_{h^\pm} +5 m^6_{h^\pm}+6 M_E^2(M_E^2-2 m^2_{h^\pm})m^2_{h^\pm} \ln\left[\frac{m^2_{h^\pm}}{M_E^2}\right]}
{12(M^2_E-m^2_{h^\pm})^4}
\label{damu}.
\end{align}

{\it Dark matter}:

Assuming the lightest Majorana particle of $N_R$ as our DM candidate, which is denoted by $X$,
we find the dominant mode to explain the observed relic density $\Omega h^2\approx0.12$~\cite{Ade:2013zuv}.
Our dominant non-relativistic cross section comes from $2 X\to  2 s$ with $t$- and $u$-channels~\footnote{Even when there is $N_{CB}$ enhancement for the processes of $\gamma\gamma$ or $\gamma Z$ final state modes, these cross sections are still subdominant.}, and 
its formula is given by
\begin{align}
\sigma v_{\rm rel}\approx 
\frac{M^6_X}{3\pi v_S^4}\sqrt{1-\frac{m^2_S}{M^2_X}}\left(41 M^4_X-38 M^2_X m^2_S+9 m^4_S\right)v^2_{\rm rel}\equiv 
 b_{\rm eff} v^2_{\rm rel}.
\end{align}
Then the relic density is formulated by
{\begin{align}
\Omega h^2\approx
\frac{1.07\times10^9 x_f^2}{3 \sqrt{g_*} M_P b_{\rm eff} },
\label{eq:relic-formula}
\end{align}
}
where $M_P\approx 1.22\times10^{19}$ GeV is the Planck mass, $g_*\approx 100$ is the total number of effective relativistic degrees of freedom at the time of freeze-out, and  $x_F\approx25$.
In our numerical analysis below, we set the allowed region to be
\begin{align}
0.11\lesssim \Omega h^2\lesssim 0.13\label{eq:relicexp},
\end{align}
where mass relation $M_X< \{ M_E, m_{h^\pm},m_{k^{\pm\pm}} \}$ is expected to stabilize DM.

{\it Numerical results}:

Now we randomly select values of the twelve parameters within the corresponding ranges
\begin{align}
& v_S \in [2\,\text{TeV}, 3\,\text{TeV}],\quad   \mu=\mu_{S} \in [0\,\text{}, 1][\text{TeV}],\quad   
M_X \in [m_S\,\text{}, v_S],\nn\\ 
&m_{k^{\pm\pm}}\in [M_X\,\text{}, 5\,\text{TeV}],\quad
M_E=M_N=m_{h^\pm} \in [M_X\,\text{}, 10\,\text{TeV}],\nn\\
& m_\ell \in [m_e\,\text{}, m_\tau\,\text{}],\quad  f=g=h \in [-1,1], 
	\label{range_scanning}
\end{align}
to reproduce the neutrino mass scale $0.001\ {\rm eV}\lesssim m_\nu \lesssim 0.1\ {\rm eV}$, the anomalous magnetic moment $2.0\times 10^{-9}\lesssim \Delta a_\mu\lesssim 4.2\times10^{-9}$ in Eq.~(\ref{dev12}), the measured relic density $0.11\lesssim \Omega h^2\lesssim 0.13$ in Eq.~(\ref{eq:relicexp}), and the decay rate to the two photons of the doubly charged bosons $k^{\pm\pm}$ observed by the 750 GeV diphoton excess 5.3 GeV $\lesssim\Gamma_S\lesssim 45$ GeV in Eq.~(\ref{eq:dip}). Here we fix $N_{CB}=[6000,9000]$, $m_S=750$ GeV is the new particle, $m_e=0.51$ MeV is the electron mass, and $m_\tau=1.776$ GeV is the tauon mass.
Then we have obtained the following constrained parameters with five millions random sampling points:
\begin{align}
N_{CB}=6000:&\nn\\
& v_S \in [2\,\text{}, 2.8\,\text{}]\ [\text{TeV}],\quad \mu_{S} \in [0.3\,\text{}, 1][\text{TeV}],\quad   
M_X \in [0.8\text{}, 1.8]\ [\text{TeV}],\nn\\ 
&m_{k^{\pm\pm}}\in [0.9\,\text{}, 2\,\text{}]\ [\text{TeV}],\quad
M_E \in [M_X\,\text{}, 6\,\text{TeV}], \quad m_{h^\pm} \in [M_X\,\text{}, 8\,\text{TeV}],\nn\\
& |f|=|g| \in [0.5,1],
\label{range-scanning6k}
\end{align}
\begin{align}
N_{CB}=9000:&\nn\\
& \mu_{S} \in [0.2\,\text{}, 1][\text{TeV}],\quad  M_X \in [0.8\text{}, 2.1]\ [\text{TeV}],\nn\\ 
& m_{k^{\pm\pm}}\in [1.0\,\text{}, 2.5\,\text{}]\ [\text{TeV}],\quad
M_E \in [M_X\,\text{}, 8\,\text{TeV}], \quad m_{h^\pm} \in [M_X\,\text{}, 9\,\text{TeV}], \nn\\
& |f|=|g| \in [0.5,1].
\label{range-scanning9k}
\end{align}
These above results suggest that $N_{CB}=9000$ gives larger number of solutions  than those of $N_{CB}=6000$, that is expected from Fig.~\ref{mk-ncb}. Also both  the allowed regions of $m_{k^{\pm\pm}}$ and $\mu_S$ directly reflect the results of this figures. 
The Yukawa couplings of $f$ and $g$ needs rather large values that are required to satisfy muon anomalous magnetic moment.
It is worth mentioning that  there exist lepton flavor violating processes (LFVs) whenever we have the contributions of the muon $g-2$ as discussed in Eq.~(\ref{eq:g-2}), although serious analysis is beyond our scope due to the very complicated neutrino sector.
These processes provide some constraints such as Yukawas ($f$ and $g$ in our case) and/or the mediating particles ($m_{h^\pm}$, and $M_E$ in our case). Even when our Yukawa couplings $f$ and $g$ are relatively large, we expect that LFVs could be suppressed by the mediating particles; $m_{h^\pm}$, $M_E$, all of which are ${\cal O}$(1) TeV.
{Here we especially show a sample point to satisfy the LFV process of  $\mu\to e\gamma$ at the one-loop level, which gives the most stringent constraint. Therefore the upper limit of the branching ratio is given by ${\rm BR}(\mu\to e\gamma)\lesssim 5.7\times10^{-13}$ from the MEG~\cite{Adam:2013mnn} at the 95 \% confidential level, and its formula is given by
\begin{align}
&{\rm BR}(\mu\to e\gamma)\approx \frac{3\alpha_{\rm em}N_{CB}^2}{32\pi {\rm G_F^2}}\left|(g^*g)_{21} F(E,h)-\frac{(f^*f)_{21}}{3 m_{h^\pm}^2}\right|^2\label{eq:lfvs}.
\end{align}
Then each of the sample point for $N_{CB}=(6000,9000)$ is given as
\begin{align}
&N_{CB}=6000:\quad {\rm BR}(\mu\to e\gamma)\approx 2.98\times10^{-13},\nn\\
& 
M_E = 1.43 [\text{TeV}], \ m_{h^\pm}  =7.35 [\text{TeV}], 
(g^*g)_{21}=0.755, \ (f^*f)_{21}=0.777,
\label{lfv-6k}\\
&N_{CB}=9000:\quad {\rm BR}(\mu\to e\gamma)\approx 4.74\times10^{-13}, \nn\\
& 
M_E = 2.25 [\text{TeV}], \ m_{h^\pm}  =6.06 [\text{TeV}], 
(g^*g)_{21}=0.631, \ (f^*f)_{21}=0.511,
\label{lfv-9k}
\end{align}
where these sample points satisfy the allowed regions in Eqs.~(\ref{range-scanning6k}) and (\ref{range-scanning9k}) respectively.
}

\begin{figure}[tb]
\begin{center}
\includegraphics[width=70mm]{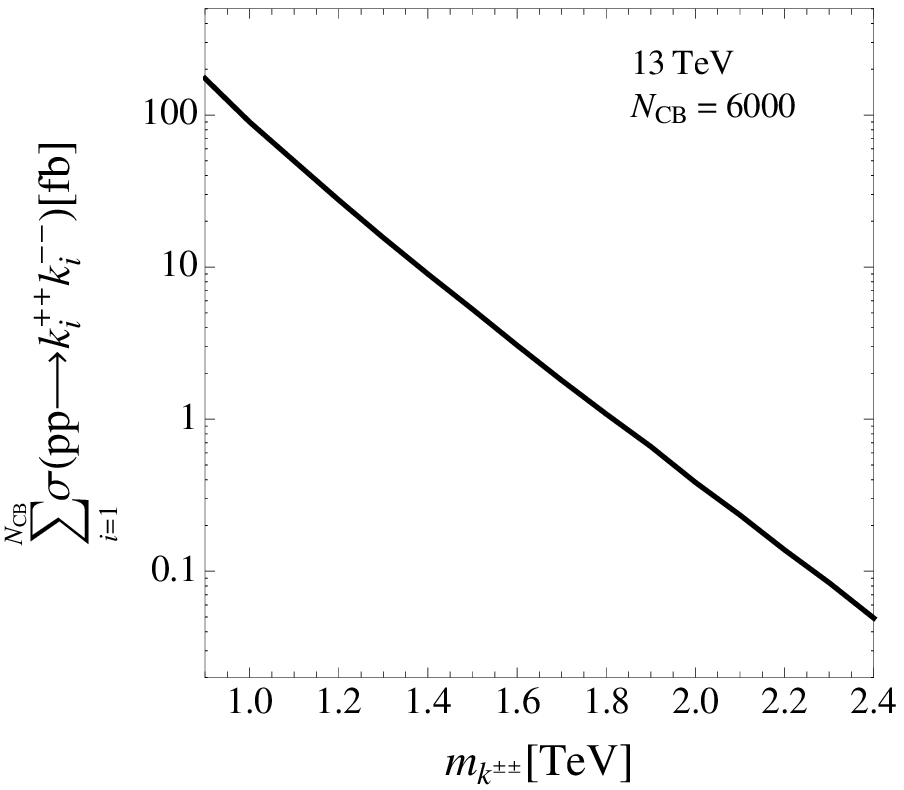}
\includegraphics[width=70mm]{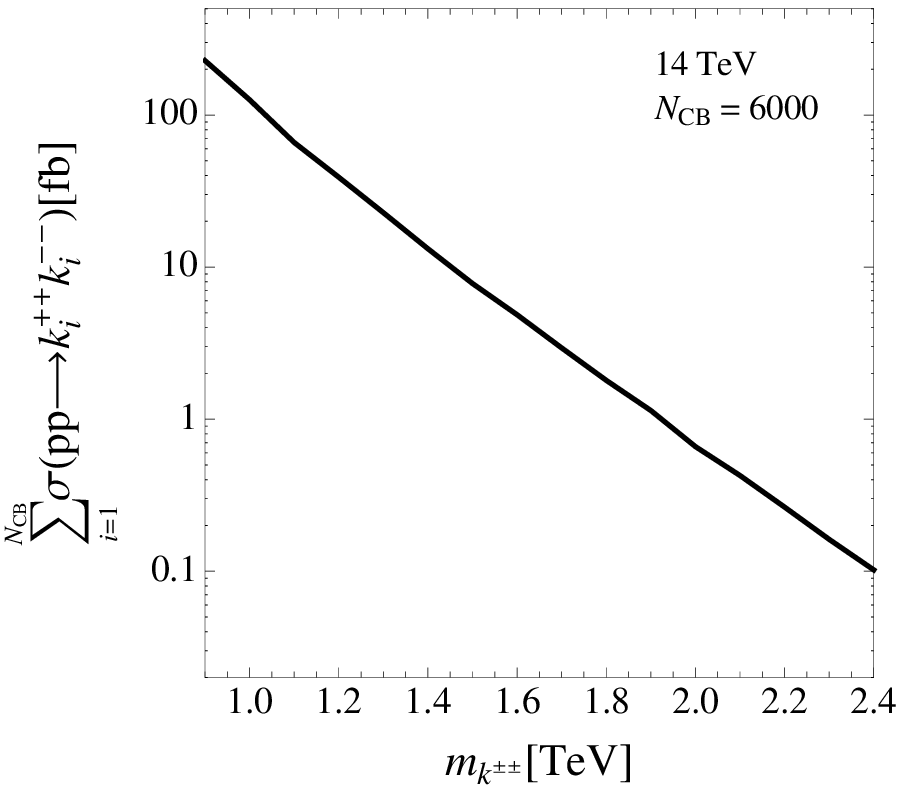}
\caption{ Sum of cross sections for doubly charged scalar production $pp \to \gamma^*/Z* \to k_i^{++} k_i^{--}$ at the LHC 13(14) TeV.
}   \label{fig:XS}
\end{center}\end{figure}
We also estimate the cross section of doubly charged scalar production, i.e. $pp \to \gamma^*/Z^* \to k^{++} k^{--}$. 
Although each pair production cross section is small the sum of the cross section for $N_{CB}$ pair can be sizable.
The production cross section is numerically estimated by CalcHEP~\cite{Belyaev:2012qa} implementing relevant interactions and using {\tt CTEQ6L} PDF~\cite{Nadolsky:2008zw}.
The left(right) plots in Fig.~\ref{fig:XS} show the sum of the $k^{++}k^{--}$ production cross section at the LHC 13(14) TeV applying $N_{CB} =6000$.
Note that the total cross section is simply $N_{CB} \times $(each $k^{++} k^{--}$ production cross section).
We thus find that the doubly charged scalar could be produced at the LHC run-II with $O(100)$ fb cross section when $m_{k^{\pm \pm}} \sim 1$ TeV.
The doubly charged scalar then decays as $k^{\pm \pm} \to h^\pm h^\pm \to \ell^\pm \ell^\pm \nu \bar \nu$ where $\ell =e, \mu$ and $\tau$.
Therefore the signal of the $k^{++}k^{--}$ pair is four charged lepton plus missing transverse energy.

\section{ Conclusions and discussions}
We have proposed a four-loop induced radiative neutrino mass model inspired by the diphoton excess at 750 GeV recently reported by ATLAS and CMS, in which a sizable diphoton excess is obtained via photon fusion introducing multi doubly-charged scalar bosons. 
The sizable neutrino mass scale has been obtained due to the enhancement of the number of doubly charged bosons $N_{CB}$. 
Also we have discussed the muon anomalous magnetic moment, and a dark matter candidate of the lightest fermion $X$, and we have found that the main process to explain the correct relic density relies on the final state of the new particle at 750 GeV through the $t$- and $u$-channels. Finally we have shown the numerical results and have obtained allowed region of several physical values in our model, as can be seen in Eqs~(\ref{range-scanning6k}) for $N_{CB}=6000$ and  Eqs~(\ref{range-scanning9k}) for $N_{CB}=9000$ respectively.
The doubly charged scalar production cross section has been numerically estimated. Then we have found that sum of the pair production cross section can be as large as $O(100)$ fb for $m_{k^{\pm \pm}} \sim 1$ TeV. 
Therefore our model could be tested at the LHC run-II by searching for the signal of four charged lepton plus missing transverse energy which is obtained as $k^{++}k^{--} \to h^+h^+h^-h^- \to \ell^+ \ell^+ \ell^- \ell^- + 4\nu$.
Further analysis of the signal is left as future work.

\section*{Acknowledgments}
\vspace{0.5cm}
H.O. thanks to Prof. Shinya Kanemura, Prof. Seong Chan Park, Dr. Kenji Nishiwaki, Dr. Yuta Orikasa, and Dr. Ryoutaro Watanabe for fruitful discussions. 
H. O. is sincerely grateful for all the KIAS members, Korean cordial persons, foods, culture, weather, and all the other things.


\begin{thebibliography}{99}

\bibitem{ATLAS-CONF-2015-081} 
  The ATLAS collaboration,
  ATLAS-CONF-2015-081.
  
\bibitem{CMS:2015dxe} 
  CMS Collaboration [CMS Collaboration],
  collisions at 13TeV,''
  CMS-PAS-EXO-15-004.
  
  
\bibitem{Harigaya:2015ezk} 
  K.~Harigaya and Y.~Nomura,
  arXiv:1512.04850 [hep-ph].
  
\bibitem{Mambrini:2015wyu} 
  Y.~Mambrini, G.~Arcadi and A.~Djouadi,
  arXiv:1512.04913 [hep-ph].

\bibitem{Backovic:2015fnp} 
  M.~Backovic, A.~Mariotti and D.~Redigolo,
  arXiv:1512.04917 [hep-ph].

\bibitem{Angelescu:2015uiz} 
  A.~Angelescu, A.~Djouadi and G.~Moreau,
  arXiv:1512.04921 [hep-ph].

\bibitem{Nakai:2015ptz} 
  Y.~Nakai, R.~Sato and K.~Tobioka,
  arXiv:1512.04924 [hep-ph].

\bibitem{Knapen:2015dap} 
  S.~Knapen, T.~Melia, M.~Papucci and K.~Zurek,
  arXiv:1512.04928 [hep-ph].

\bibitem{Buttazzo:2015txu} 
  D.~Buttazzo, A.~Greljo and D.~Marzocca,
  arXiv:1512.04929 [hep-ph].

\bibitem{Pilaftsis:2015ycr} 
  A.~Pilaftsis,
  arXiv:1512.04931 [hep-ph].
  
\bibitem{Franceschini:2015kwy} 
  R.~Franceschini {\it et al.},
  arXiv:1512.04933 [hep-ph].

\bibitem{DiChiara:2015vdm} 
  S.~Di Chiara, L.~Marzola and M.~Raidal,
  arXiv:1512.04939 [hep-ph].

\bibitem{Higaki:2015jag} 
  T.~Higaki, K.~S.~Jeong, N.~Kitajima and F.~Takahashi,
  arXiv:1512.05295 [hep-ph].

\bibitem{McDermott:2015sck} 
  S.~D.~McDermott, P.~Meade and H.~Ramani,
  arXiv:1512.05326 [hep-ph].

\bibitem{Ellis:2015oso} 
  J.~Ellis, S.~A.~R.~Ellis, J.~Quevillon, V.~Sanz and T.~You,
  arXiv:1512.05327 [hep-ph].

\bibitem{Low:2015qep} 
  M.~Low, A.~Tesi and L.~T.~Wang,
  arXiv:1512.05328 [hep-ph].

\bibitem{Bellazzini:2015nxw} 
  B.~Bellazzini, R.~Franceschini, F.~Sala and J.~Serra,
  arXiv:1512.05330 [hep-ph].

\bibitem{Gupta:2015zzs} 
  R.~S.~Gupta, S.~Jager, Y.~Kats, G.~Perez and E.~Stamou,
  arXiv:1512.05332 [hep-ph].
  
\bibitem{Petersson:2015mkr} 
  C.~Petersson and R.~Torre,
  arXiv:1512.05333 [hep-ph].

\bibitem{Molinaro:2015cwg} 
  E.~Molinaro, F.~Sannino and N.~Vignaroli,
  arXiv:1512.05334 [hep-ph].
  
\bibitem{Dutta:2015wqh} 
  B.~Dutta, Y.~Gao, T.~Ghosh, I.~Gogoladze and T.~Li,
  arXiv:1512.05439 [hep-ph].
  
\bibitem{Cao:2015pto} 
  Q.~H.~Cao, Y.~Liu, K.~P.~Xie, B.~Yan and D.~M.~Zhang,
  arXiv:1512.05542 [hep-ph].

\bibitem{Matsuzaki:2015che} 
  S.~Matsuzaki and K.~Yamawaki,
  arXiv:1512.05564 [hep-ph].

\bibitem{Kobakhidze:2015ldh} 
  A.~Kobakhidze, F.~Wang, L.~Wu, J.~M.~Yang and M.~Zhang,
  arXiv:1512.05585 [hep-ph].

\bibitem{Martinez:2015kmn} 
  R.~Martinez, F.~Ochoa and C.~F.~Sierra,
  arXiv:1512.05617 [hep-ph].

\bibitem{Cox:2015ckc} 
  P.~Cox, A.~D.~Medina, T.~S.~Ray and A.~Spray,
  arXiv:1512.05618 [hep-ph].
  
\bibitem{Becirevic:2015fmu} 
  D.~Becirevic, E.~Bertuzzo, O.~Sumensari and R.~Z.~Funchal,
  arXiv:1512.05623 [hep-ph].

\bibitem{No:2015bsn} 
  J.~M.~No, V.~Sanz and J.~Setford,
  arXiv:1512.05700 [hep-ph].
  
\bibitem{Demidov:2015zqn} 
  S.~V.~Demidov and D.~S.~Gorbunov,
  arXiv:1512.05723 [hep-ph].

\bibitem{Chao:2015ttq} 
  W.~Chao, R.~Huo and J.~H.~Yu,
  arXiv:1512.05738 [hep-ph].
  
\bibitem{Fichet:2015vvy} 
  S.~Fichet, G.~von Gersdorff and C.~Royon,
  arXiv:1512.05751 [hep-ph].

\bibitem{Curtin:2015jcv} 
  D.~Curtin and C.~B.~Verhaaren,
  arXiv:1512.05753 [hep-ph].

\bibitem{Bian:2015kjt} 
  L.~Bian, N.~Chen, D.~Liu and J.~Shu,
  arXiv:1512.05759 [hep-ph].

\bibitem{Chakrabortty:2015hff} 
  J.~Chakrabortty, A.~Choudhury, P.~Ghosh, S.~Mondal and T.~Srivastava,
  arXiv:1512.05767 [hep-ph].

\bibitem{Ahmed:2015uqt} 
  A.~Ahmed, B.~M.~Dillon, B.~Grzadkowski, J.~F.~Gunion and Y.~Jiang,
  arXiv:1512.05771 [hep-ph].
  
\bibitem{Agrawal:2015dbf} 
  P.~Agrawal, J.~Fan, B.~Heidenreich, M.~Reece and M.~Strassler,
  arXiv:1512.05775 [hep-ph].

\bibitem{Csaki:2015vek} 
  C.~Csaki, J.~Hubisz and J.~Terning,
  arXiv:1512.05776 [hep-ph].

\bibitem{Falkowski:2015swt} 
  A.~Falkowski, O.~Slone and T.~Volansky,
  arXiv:1512.05777 [hep-ph].

\bibitem{Aloni:2015mxa} 
  D.~Aloni, K.~Blum, A.~Dery, A.~Efrati and Y.~Nir,
  arXiv:1512.05778 [hep-ph].
  
\bibitem{Bai:2015nbs} 
  Y.~Bai, J.~Berger and R.~Lu,
  arXiv:1512.05779 [hep-ph].

\bibitem{Gabrielli:2015dhk} 
  E.~Gabrielli, K.~Kannike, B.~Mele, M.~Raidal, C.~Spethmann and H.~Veermae,
  arXiv:1512.05961 [hep-ph].

\bibitem{Benbrik:2015fyz} 
  R.~Benbrik, C.~H.~Chen and T.~Nomura,
  arXiv:1512.06028 [hep-ph].

\bibitem{Kim:2015ron} 
  J.~S.~Kim, J.~Reuter, K.~Rolbiecki and R.~R.~de Austri,
  arXiv:1512.06083 [hep-ph].

\bibitem{Alves:2015jgx} 
  A.~Alves, A.~G.~Dias and K.~Sinha,
  arXiv:1512.06091 [hep-ph].

\bibitem{Megias:2015ory} 
  E.~Megias, O.~Pujolas and M.~Quiros,
  arXiv:1512.06106 [hep-ph].

\bibitem{Carpenter:2015ucu} 
  L.~M.~Carpenter, R.~Colburn and J.~Goodman,
  arXiv:1512.06107 [hep-ph].

\bibitem{Bernon:2015abk} 
  J.~Bernon and C.~Smith,
  arXiv:1512.06113 [hep-ph].

\bibitem{Chao:2015nsm} 
  W.~Chao,
  arXiv:1512.06297 [hep-ph].

\bibitem{Arun:2015ubr} 
  M.~T.~Arun and P.~Saha,
  arXiv:1512.06335 [hep-ph].

\bibitem{Han:2015cty} 
  C.~Han, H.~M.~Lee, M.~Park and V.~Sanz,
  arXiv:1512.06376 [hep-ph].

\bibitem{Chang:2015bzc} 
  S.~Chang,
  arXiv:1512.06426 [hep-ph].

\bibitem{Chakraborty:2015jvs} 
  I.~Chakraborty and A.~Kundu,
  arXiv:1512.06508 [hep-ph].

\bibitem{Ding:2015rxx} 
  R.~Ding, L.~Huang, T.~Li and B.~Zhu,
  arXiv:1512.06560 [hep-ph].
  
\bibitem{Han:2015dlp} 
  H.~Han, S.~Wang and S.~Zheng,
  arXiv:1512.06562 [hep-ph].

\bibitem{Han:2015qqj} 
  X.~F.~Han and L.~Wang,
  arXiv:1512.06587 [hep-ph].

\bibitem{Luo:2015yio} 
  M.~x.~Luo, K.~Wang, T.~Xu, L.~Zhang and G.~Zhu,
  arXiv:1512.06670 [hep-ph].

\bibitem{Chang:2015sdy} 
  J.~Chang, K.~Cheung and C.~T.~Lu,
  arXiv:1512.06671 [hep-ph].

\bibitem{Bardhan:2015hcr} 
  D.~Bardhan, D.~Bhatia, A.~Chakraborty, U.~Maitra, S.~Raychaudhuri and T.~Samui,
  arXiv:1512.06674 [hep-ph].
  
\bibitem{Feng:2015wil} 
  T.~F.~Feng, X.~Q.~Li, H.~B.~Zhang and S.~M.~Zhao,
  arXiv:1512.06696 [hep-ph].

\bibitem{Antipin:2015kgh} 
  O.~Antipin, M.~Mojaza and F.~Sannino,
  arXiv:1512.06708 [hep-ph].

\bibitem{Wang:2015kuj} 
  F.~Wang, L.~Wu, J.~M.~Yang and M.~Zhang,
  arXiv:1512.06715 [hep-ph].
  
\bibitem{Cao:2015twy} 
  J.~Cao, C.~Han, L.~Shang, W.~Su, J.~M.~Yang and Y.~Zhang,
  arXiv:1512.06728 [hep-ph].

\bibitem{Huang:2015evq} 
  F.~P.~Huang, C.~S.~Li, Z.~L.~Liu and Y.~Wang,
  arXiv:1512.06732 [hep-ph].

\bibitem{Liao:2015tow} 
  W.~Liao and H.~q.~Zheng,
  arXiv:1512.06741 [hep-ph].
  
\bibitem{Heckman:2015kqk} 
  J.~J.~Heckman,
  arXiv:1512.06773 [hep-ph].

\bibitem{Dhuria:2015ufo} 
  M.~Dhuria and G.~Goswami,
  arXiv:1512.06782 [hep-ph].

\bibitem{Bi:2015uqd} 
  X.~J.~Bi, Q.~F.~Xiang, P.~F.~Yin and Z.~H.~Yu,
  arXiv:1512.06787 [hep-ph].

\bibitem{Kim:2015ksf} 
  J.~S.~Kim, K.~Rolbiecki and R.~R.~de Austri,
  arXiv:1512.06797 [hep-ph].
  
\bibitem{Berthier:2015vbb} 
  L.~Berthier, J.~M.~Cline, W.~Shepherd and M.~Trott,
  arXiv:1512.06799 [hep-ph].

\bibitem{Cho:2015nxy} 
  W.~S.~Cho, D.~Kim, K.~Kong, S.~H.~Lim, K.~T.~Matchev, J.~C.~Park and M.~Park,
  arXiv:1512.06824 [hep-ph].

\bibitem{Cline:2015msi} 
  J.~M.~Cline and Z.~Liu,
  arXiv:1512.06827 [hep-ph].
  
\bibitem{Bauer:2015boy} 
  M.~Bauer and M.~Neubert,
  arXiv:1512.06828 [hep-ph].

\bibitem{Chala:2015cev} 
  M.~Chala, M.~Duerr, F.~Kahlhoefer and K.~Schmidt-Hoberg,
  arXiv:1512.06833 [hep-ph].

\bibitem{Barducci:2015gtd} 
  D.~Barducci, A.~Goudelis, S.~Kulkarni and D.~Sengupta,
  arXiv:1512.06842 [hep-ph].
  
\bibitem{Boucenna:2015pav} 
  S.~M.~Boucenna, S.~Morisi and A.~Vicente,
  arXiv:1512.06878 [hep-ph].

\bibitem{Murphy:2015kag} 
  C.~W.~Murphy,
  arXiv:1512.06976 [hep-ph].

\bibitem{Hernandez:2015ywg} 
  A.~E.~C.~Hernandez and I.~Nisandzic,
  arXiv:1512.07165 [hep-ph].

\bibitem{Dey:2015bur} 
  U.~K.~Dey, S.~Mohanty and G.~Tomar,
  arXiv:1512.07212 [hep-ph].

\bibitem{Pelaggi:2015knk} 
  G.~M.~Pelaggi, A.~Strumia and E.~Vigiani,
  arXiv:1512.07225 [hep-ph].

\bibitem{deBlas:2015hlv} 
  J.~de Blas, J.~Santiago and R.~Vega-Morales,
  arXiv:1512.07229 [hep-ph].

\bibitem{Belyaev:2015hgo} 
  A.~Belyaev, G.~Cacciapaglia, H.~Cai, T.~Flacke, A.~Parolini and H.~Serodio,
  arXiv:1512.07242 [hep-ph].

\bibitem{Dev:2015isx} 
  P.~S.~B.~Dev and D.~Teresi,
  arXiv:1512.07243 [hep-ph].

\bibitem{Huang:2015rkj} 
  W.~C.~Huang, Y.~L.~S.~Tsai and T.~C.~Yuan,
  arXiv:1512.07268 [hep-ph].

\bibitem{Moretti:2015pbj} 
  S.~Moretti and K.~Yagyu,
  arXiv:1512.07462 [hep-ph].
  
\bibitem{Patel:2015ulo} 
  K.~M.~Patel and P.~Sharma,
  arXiv:1512.07468 [hep-ph].
  
\bibitem{Badziak:2015zez} 
  M.~Badziak,
  arXiv:1512.07497 [hep-ph].
  
\bibitem{Chakraborty:2015gyj} 
  S.~Chakraborty, A.~Chakraborty and S.~Raychaudhuri,
  arXiv:1512.07527 [hep-ph].

\bibitem{Cao:2015xjz} 
  Q.~H.~Cao, S.~L.~Chen and P.~H.~Gu,
  arXiv:1512.07541 [hep-ph].

\bibitem{Altmannshofer:2015xfo} 
  W.~Altmannshofer, J.~Galloway, S.~Gori, A.~L.~Kagan, A.~Martin and J.~Zupan,
  arXiv:1512.07616 [hep-ph].

\bibitem{Cvetic:2015vit} 
  M.~Cvetic, J.~Halverson and P.~Langacker,
  arXiv:1512.07622 [hep-ph].

\bibitem{Gu:2015lxj} 
  J.~Gu and Z.~Liu,
  arXiv:1512.07624 [hep-ph].

\bibitem{Allanach:2015ixl} 
  B.~C.~Allanach, P.~S.~B.~Dev, S.~A.~Renner and K.~Sakurai,
  arXiv:1512.07645 [hep-ph].

\bibitem{Davoudiasl:2015cuo} 
  H.~Davoudiasl and C.~Zhang,
  arXiv:1512.07672 [hep-ph].

\bibitem{Craig:2015lra} 
  N.~Craig, P.~Draper, C.~Kilic and S.~Thomas,
  arXiv:1512.07733 [hep-ph].

\bibitem{Das:2015enc} 
  K.~Das and S.~K.~Rai,
  arXiv:1512.07789 [hep-ph].

\bibitem{Cheung:2015cug} 
  K.~Cheung, P.~Ko, J.~S.~Lee, J.~Park and P.~Y.~Tseng,
  arXiv:1512.07853 [hep-ph].

\bibitem{Liu:2015yec} 
  J.~Liu, X.~P.~Wang and W.~Xue,
  arXiv:1512.07885 [hep-ph].
  
\bibitem{Zhang:2015uuo} 
  J.~Zhang and S.~Zhou,
  arXiv:1512.07889 [hep-ph].

\bibitem{Casas:2015blx} 
  J.~A.~Casas, J.~R.~Espinosa and J.~M.~Moreno,
  arXiv:1512.07895 [hep-ph].
  
\bibitem{Hall:2015xds} 
  L.~J.~Hall, K.~Harigaya and Y.~Nomura,
  arXiv:1512.07904 [hep-ph].
  
\bibitem{Han:2015yjk} 
  H.~Han, S.~Wang and S.~Zheng,
  arXiv:1512.07992 [hep-ph].

\bibitem{Park:2015ysf} 
  J.~C.~Park and S.~C.~Park,
  arXiv:1512.08117 [hep-ph].
  
\bibitem{Salvio:2015jgu} 
  A.~Salvio and A.~Mazumdar,
  arXiv:1512.08184 [hep-ph].

\bibitem{Chway:2015lzg} 
  D.~Chway, R.~Dermisek, T.~H.~Jung and H.~D.~Kim,
  arXiv:1512.08221 [hep-ph].
  
\bibitem{Li:2015jwd} 
  G.~Li, Y.~n.~Mao, Y.~L.~Tang, C.~Zhang, Y.~Zhou and S.~h.~Zhu,
  arXiv:1512.08255 [hep-ph].

\bibitem{Son:2015vfl} 
  M.~Son and A.~Urbano,
  arXiv:1512.08307 [hep-ph].
  
\bibitem{Tang:2015eko} 
  Y.~L.~Tang and S.~h.~Zhu,
  arXiv:1512.08323 [hep-ph].

\bibitem{An:2015cgp} 
  H.~An, C.~Cheung and Y.~Zhang,
  arXiv:1512.08378 [hep-ph].
  
\bibitem{Cao:2015apa} 
  J.~Cao, F.~Wang and Y.~Zhang,
  arXiv:1512.08392 [hep-ph].

\bibitem{Wang:2015omi} 
  F.~Wang, W.~Wang, L.~Wu, J.~M.~Yang and M.~Zhang,
  arXiv:1512.08434 [hep-ph].

\bibitem{Cai:2015hzc} 
  C.~Cai, Z.~H.~Yu and H.~H.~Zhang,
  arXiv:1512.08440 [hep-ph].

\bibitem{Cao:2015scs} 
  Q.~H.~Cao, Y.~Liu, K.~P.~Xie, B.~Yan and D.~M.~Zhang,
  arXiv:1512.08441 [hep-ph].

\bibitem{Kim:2015xyn} 
  J.~E.~Kim,
  arXiv:1512.08467 [hep-ph].
  
\bibitem{Gao:2015igz} 
  J.~Gao, H.~Zhang and H.~X.~Zhu,
  arXiv:1512.08478 [hep-ph].

\bibitem{Chao:2015nac} 
  W.~Chao,
  arXiv:1512.08484 [hep-ph].

\bibitem{Bi:2015lcf} 
  X.~J.~Bi {\it et al.},
  arXiv:1512.08497 [hep-ph].

\bibitem{Goertz:2015nkp} 
  F.~Goertz, J.~F.~Kamenik, A.~Katz and M.~Nardecchia,
  arXiv:1512.08500 [hep-ph].

\bibitem{Anchordoqui:2015jxc} 
  L.~A.~Anchordoqui, I.~Antoniadis, H.~Goldberg, X.~Huang, D.~Lust and T.~R.~Taylor,
  arXiv:1512.08502 [hep-ph].

\bibitem{Dev:2015vjd} 
  P.~S.~B.~Dev, R.~N.~Mohapatra and Y.~Zhang,
  arXiv:1512.08507 [hep-ph].

\bibitem{Bizot:2015qqo} 
  N.~Bizot, S.~Davidson, M.~Frigerio and J.-L.~Kneur,
  arXiv:1512.08508 [hep-ph].
 
\bibitem{Ibanez:2015uok} 
  L.~E.~Ibanez and V.~Martin-Lozano,
  arXiv:1512.08777 [hep-ph].
  
\bibitem{Chiang:2015tqz} 
  C.~W.~Chiang, M.~Ibe and T.~T.~Yanagida,
  arXiv:1512.08895 [hep-ph].
  
\bibitem{Kang:2015roj} 
  S.~K.~Kang and J.~Song,
  arXiv:1512.08963 [hep-ph].

\bibitem{Hamada:2015skp} 
  Y.~Hamada, T.~Noumi, S.~Sun and G.~Shiu,
  arXiv:1512.08984 [hep-ph].
  
\bibitem{Huang:2015svl} 
  X.~J.~Huang, W.~H.~Zhang and Y.~F.~Zhou,
  arXiv:1512.08992 [hep-ph].

  
\bibitem{Kanemura:2015bli} 
  S.~Kanemura, K.~Nishiwaki, H.~Okada, Y.~Orikasa, S.~C.~Park and R.~Watanabe,
  arXiv:1512.09048 [hep-ph].
  
\bibitem{Kanemura:2015vcb} 
  S.~Kanemura, N.~Machida, S.~Odori and T.~Shindou,
  arXiv:1512.09053 [hep-ph].

  
\bibitem{Low:2015qho} 
  I.~Low and J.~Lykken,
  arXiv:1512.09089 [hep-ph].

  
\bibitem{Hernandez:2015hrt} 
  A.~E.~C.~Hernandez,
  arXiv:1512.09092 [hep-ph].

  
\bibitem{Jiang:2015oms} 
  Y.~Jiang, Y.~Y.~Li and T.~Liu,
  arXiv:1512.09127 [hep-ph].
  
\bibitem{Kaneta:2015qpf} 
  K.~Kaneta, S.~Kang and H.~S.~Lee,
  arXiv:1512.09129 [hep-ph].
  
\bibitem{Marzola:2015xbh} 
  L.~Marzola, A.~Racioppi, M.~Raidal, F.~R.~Urban and H.~Veermae,
  arXiv:1512.09136 [hep-ph].



\bibitem{Ma:2015xmf} 
  E.~Ma,
  arXiv:1512.09159 [hep-ph].

\bibitem{Dasgupta:2015pbr} 
  A.~Dasgupta, M.~Mitra and D.~Borah,
  arXiv:1512.09202 [hep-ph].

  
  
\bibitem{Krauss:2002px}
  L.~M.~Krauss, S.~Nasri and M.~Trodden,
  Phys.\ Rev.\  D {\bf 67}, 085002 (2003)
  [arXiv:hep-ph/0210389].
  
  
\bibitem{Aoki:2008av}
  M.~Aoki, S.~Kanemura and O.~Seto,
  Phys.\ Rev.\ Lett.\  {\bf 102}, 051805 (2009)
  [arXiv:0807.0361].

\bibitem{Gustafsson:2012vj} 
M.~Gustafsson, J.~M.~No and M.~A.~Rivera,
  Phys.\ Rev.\ Lett.\  {\bf 110}, 211802 (2013)
arXiv:1212.4806 [hep-ph].


  
\bibitem{Aad:2015mna} 
  G.~Aad {\it et al.} [ATLAS Collaboration],
  Phys.\ Rev.\ D {\bf 92}, no. 3, 032004 (2015)
  doi:10.1103/PhysRevD.92.032004
  [arXiv:1504.05511 [hep-ex]].

\bibitem{CMS:2014onr} 
  CMS Collaboration [CMS Collaboration],
  CMS-PAS-HIG-14-006.
  
  
  
  arXiv:1512.05776 [hep-ph].
  

  

\bibitem{zee-babu}
 A.~Zee,
 Nucl.\ Phys.\ B {\bf 264}, 99  (1986); 
 K.~S.~Babu,
 Phys.\ Lett.\ B {\bf 203}, 132  (1988). 

  \bibitem{pdf} 
  K.A. Olive et al. (Particle Data Group), Chin. Phys. C, 38, 090001 (2014).


\bibitem{discrepancy1}
  F.~Jegerlehner and A.~Nyffeler,
  Phys.\ Rept.\  {\bf 477}, 1 (2009)
  [arXiv:0902.3360 [hep-ph]].
  
\bibitem{discrepancy2} 
  M.~Benayoun, P.~David, L.~Delbuono and F.~Jegerlehner,
  Eur.\ Phys.\ J.\ C {\bf 72}, 1848 (2012)
 [arXiv:1106.1315 [hep-ph]].



\bibitem{Ade:2013zuv} 
  P.~A.~R.~Ade {\it et al.}  [Planck Collaboration],
  Astron.\ Astrophys.\  (2014)
  [arXiv:1303.5076 [astro-ph.CO]].

\bibitem{Adam:2013mnn} 
  J.~Adam {\it et al.} [MEG Collaboration],
  Phys.\ Rev.\ Lett.\  {\bf 110}, 201801 (2013)
  [arXiv:1303.0754 [hep-ex]].
  


\bibitem{Belyaev:2012qa} 
  A.~Belyaev, N.~D.~Christensen and A.~Pukhov,
  Comput.\ Phys.\ Commun.\  {\bf 184}, 1729 (2013)
  [arXiv:1207.6082 [hep-ph]].
  
\bibitem{Nadolsky:2008zw} 
  P.~M.~Nadolsky, H.~L.~Lai, Q.~H.~Cao, J.~Huston, J.~Pumplin, D.~Stump, W.~K.~Tung and C.-P.~Yuan,
  Phys.\ Rev.\ D {\bf 78}, 013004 (2008)
  [arXiv:0802.0007 [hep-ph]].
  
  
\end{thebibliography}
\end{document}